\documentclass{aa}

\usepackage[english]{babel}
\usepackage{graphicx}
\usepackage[varg]{txfonts}
\usepackage{natbib}
\usepackage{longtable}
\usepackage{array}
\usepackage{multirow}

\bibpunct{(}{)}{;}{a}{}{,}
\begin{document}

\title{Rotational spectroscopic study and astronomical search for propiolamide in Sgr B2(N)
\thanks{Table 2 is only available in electronic form
at the CDS via anonymous ftp to cdsarc.u-strasbg.fr (130.79.128.5) or via http://cdsweb.u-strasbg.fr/cgi-bin/qcat?J/A+A/
}}

   \author{E. R. Alonso\inst{1,2}
   \and L. Kolesnikov\'{a}\inst{3}
   \and A. Belloche\inst{4}
   \and S. Mata\inst{5}
   \and R. T. Garrod\inst{6}
   \and A. Jabri\inst{5}
   \and I. Le\'{o}n\inst{5}
   \and J.-C. Guillemin\inst{7}
   \and H.~S.~P.~M\"uller\inst{8}
   \and K. M. Menten\inst{4}
    \and J. L. Alonso\inst{5}
          }

\institute{Instituto Biofisika (UPV/EHU, CSIC), University of the Basque Country, Leioa, Spain
\and
Fundación Biofísica Bizkaia / Biofisika Bizkaia Fundazioa (FBB), Barrio Sarriena s/n, Leioa, Spain
\and
Department of Analytical Chemistry, University of Chemistry and Technology,
Technick\'{a} 5, 166 28 Prague 6, Czech Republic\\\email{lucie.kolesnikova@vscht.cz}
\and
Max-Planck-Institut für Radioastronomie, Auf dem Hügel 69, 53121 Bonn, Germany
\and Grupo de Espectroscopia Molecular (GEM), Edificio Quifima, Área de Química-Física, Laboratorios de Espectroscopia y Bioespectroscopia, Parque Científico UVa, Unidad Asociada CSIC, Universidad de Valladolid, 47011 Valladolid, Spain
\and
Departments of Chemistry and Astronomy, University of Virginia, Charlottesville, VA 22904, USA
\and
Univ Rennes, Ecole Nationale Sup\'{e}rieure de Chimie de Rennes, CNRS, ISCR – UMR 6226, F-35000 Rennes, France
\and
I. Physikalisches Institut, Universit{\"a}t zu K{\"o}ln, Z{\"u}lpicher Str. 77, 50937 K{\"o}ln, Germany}

\date{Received ; accepted }

\titlerunning{The rotational spectrum of propiolamide}
\authorrunning{Alonso et al.}


\abstract
   {For all the amides detected in the interstellar medium (ISM), the corresponding nitriles or isonitriles have also been detected in the ISM, some of which have relatively high abundances. Among the abundant nitriles for which the corresponding amide has not yet been detected is cyanoacetylene (HCCCN), whose amide counterpart is propiolamide (HCCC(O)NH$_{2}$).}
   {With the aim of supporting searches for this amide in the ISM, we provide a complete rotational study of propiolamide from 6 GHz to 440 GHz.
      }
   {Time-domain Fourier transform microwave (FTMW) spectroscopy under supersonic expansion conditions between 6 GHz and 18 GHz was used to accurately measure and analyze ground-state rotational transitions with resolved hyperfine structure arising from nuclear quadrupole coupling interactions of the $^{14}$N nucleus. We combined this technique with the frequency-domain room-temperature millimeter wave and submillimeter wave spectroscopies from 75 GHz to 440 GHz in order to record and assign the rotational spectra in the ground state and in the low-lying excited vibrational states. We used the ReMoCA spectral line survey performed with the Atacama Large Millimeter/submillimeter Array toward the star-forming region Sgr B2(N) to search for propiolamide.}
   {We identified and measured more than 5500 distinct frequency lines of propiolamide in the laboratory. These lines were fitted using an effective semi-rigid rotor Hamiltonian with nuclear quadrupole coupling interactions taken into consideration. We obtained accurate sets of spectroscopic parameters for the ground state and the three low-lying excited vibrational states. We report the nondetection of propiolamide toward the hot cores Sgr~B2(N1S) and Sgr~B2(N2). We find that propiolamide is at least 50 and 13 times less abundant than acetamide in Sgr~B2(N1S) and Sgr~B2(N2), respectively, indicating that the abundance difference between both amides is more pronounced by at least a factor of 8 and 2, respectively, than for their corresponding nitriles.}
   {Although propiolamide has yet to be included in astrochemical modeling
networks, the observed upper limit to the ratio of propiolamide to
acetamide seems consistent with the ratios of related species as
determined from past simulations. The comprehensive spectroscopic data presented in this paper will aid future astronomical searches.}

   \keywords{astrochemistry – ISM: molecules – line: identification – ISM: individual objects: Sagittarius B2 -- astronomical databases: miscellaneous
}

   \maketitle
%

\section{Introduction}
\label{sect_intro}

\begin{figure}[ht]
   \centering
   \includegraphics[width=6.0cm]{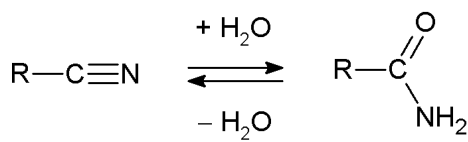}
      \caption{Schematic representation of the hydration of nitriles to amides and the dehydration of amides to nitriles.}\label{scheme}
   \end{figure}

The discovery of molecules in the interstellar medium (ISM) is greatly facilitated by our ability to predict which compounds are present in that environment. Many interstellar molecules form in the icy mantles of dust grains, including complex organic molecules (COMs\footnote{COMs are defined to contain six or more atoms; at least one of these atoms is carbon.}; \cite{Herbst2009}). Laboratory simulations of the chemistry of grains in the ISM could provide lists of target compounds.
For example, it has recently been shown that, under hydrogen radical bombardments at 10 K, the aldehydes are not reduced into the corresponding alcohols \citep{Jonusas2017} except for formaldehyde which gives methanol
\citep{Hiraoka1994,Watanabe2002,Fuchs2009,Pirim2011}.
Based on similar experiments, nitriles do not lead to amines \citep{Krim2019} except for HCN, which provides methylamine in a very low yield \citep{Theule2011}. On the other hand, to the best of our knowledge, the relation between nitriles and amides by the hydration of the former or the dehydration of the latter (see Fig. \ref{scheme}) has never been studied in such laboratory simulations. We can however observe that some couples of nitrile and amide are present in the list of compounds detected in the ISM, such as hydrogen cyanide (HCN) and formamide (NH$_2$CHO) or acetonitrile (CH$_3$CN) and acetamide (CH$_3$C(O)NH$_2$).
However, such comparisons are difficult because for most of the amides corresponding to the nitriles observed in the ISM, one of the most powerful tools for such detections, the millimeter wave spectrum, has never been recorded. Only microwave spectra can be found for propionamide \citep{Marstokk1996}, acrylamide \citep{Marstokk2000}, or propiolamide \citep{Little1978}. The particular case of glycinamide (NH$_{2}$CH$_{2}$C(O)NH$_{2}$) has to be detailed. Many attempts have tried to detect amino acids in the ISM and particularly the simplest one, glycine, but only a potential precursor of glycine, aminoacetonitrile (NH$_{2}$CH$_{2}$CN), has been detected \citep{Belloche2008}. In the hypothesis that this compound could be a precursor of glycine by hydrolysis, the addition of one molecule of water would give the glycinamide intermediate. The microwave spectrum of this compound was recently published \citep{Alonso2018}, but the millimeter wave spectrum is still missing.
On the other hand, this is not the case of glycolamide, a glycine isomer, whose millimeter wave spectrum was recorded very recently \citep{Sanz-Novo2020}.

To support the hypothesis of a correlation between the presence of a nitrile and its corresponding amide, we first looked at which nitriles are the most abundant in the ISM. Unambiguously, after hydrogen cyanide (HCN) and acetonitrile (CH$_{3}$CN), propionitrile (C$_2$H$_{5}$CN) and cyanoacetylene (HC$_3$N) are the most abundant derivatives \citep{McGuire2018}. HC$_3$N is ubiquitous in the Universe, having been detected in comets, the atmosphere of Titan, and in many places of the ISM. Therefore, propiolamide (HCCC(O)NH$_{2}$) is plausible interstellar candidate. In addition, this eight-atom molecule has fewer atoms than the detected acetamide and contains a carbon-carbon triple bond, a widely represented functional group among the known circumstellar and interstellar compounds such as cyanopolyynes, alkynes, alkyne radicals or anions, magnesium acetylide (MgCCH), propynal (HCCCHO), ethynyl isocyanide (HCCNC), oxo-penta-2,4-diynyl (HC$_{5}$O), and oxo-hepta-2,4,6-triynyl (HC$_{7}$O) \citep{McGuire2018}.

Since reliable interstellar searches for propiolamide should be based on rotational transitions measured directly in the laboratory or
those predicted from a data set that covers a broad spectral range,
in this work we provide a detailed rotational study of this amide up to 440 GHz. Two different high-resolution spectroscopic techniques were employed for this purpose. Narrowband cavity-based Fourier transform microwave (FTMW) spectroscopy was used to measure the jet-cooled spectrum of propiolamide up to 18 GHz and to analyze the $^{14}$N nuclear quadrupole hyperfine structure of the ground-state rotational transitions. Millimeter and submillimeter wave spectroscopic techniques were used to record the room-temperature spectrum up to 480 GHz and to identify higher-$J$ ground-state transitions and the pure rotational spectra in excited vibrational states. The present work significantly extends the knowledge of the rotational spectrum of propiolamide and provides sufficiently precise laboratory information to search for this amide in space. We chose as a target for this interstellar search the high-mass star-forming region Sagittarius (Sgr) B2(N), which is located close to the Galactic center. We use our latest imaging spectral line survey performed toward this source with the Atacama Large Millimeter/submillimeter Array (ALMA) in the 3~mm atmospheric window, the ReMoCA survey, which recently led to the identification of urea \citep[][]{Belloche19}.

The article is structured as follows. Details about the laboratory
experiments are given in Sect.~\ref{s:experiments} and the analysis of the
recorded spectra of propiolamide is described in Sect.~\ref{s:analysis}.
The search for propiolamide toward Sgr~B2(N) is reported in
Sect.~\ref{s:astro}. We discuss our results in Sect.~\ref{s:discussion} and
give our conclusions in Sect.~\ref{s:conclusions}.

      \begin{figure*}[ht]
   \centering
   \includegraphics[trim = 28mm 0mm 20mm 0mm, clip, width=16.5cm]{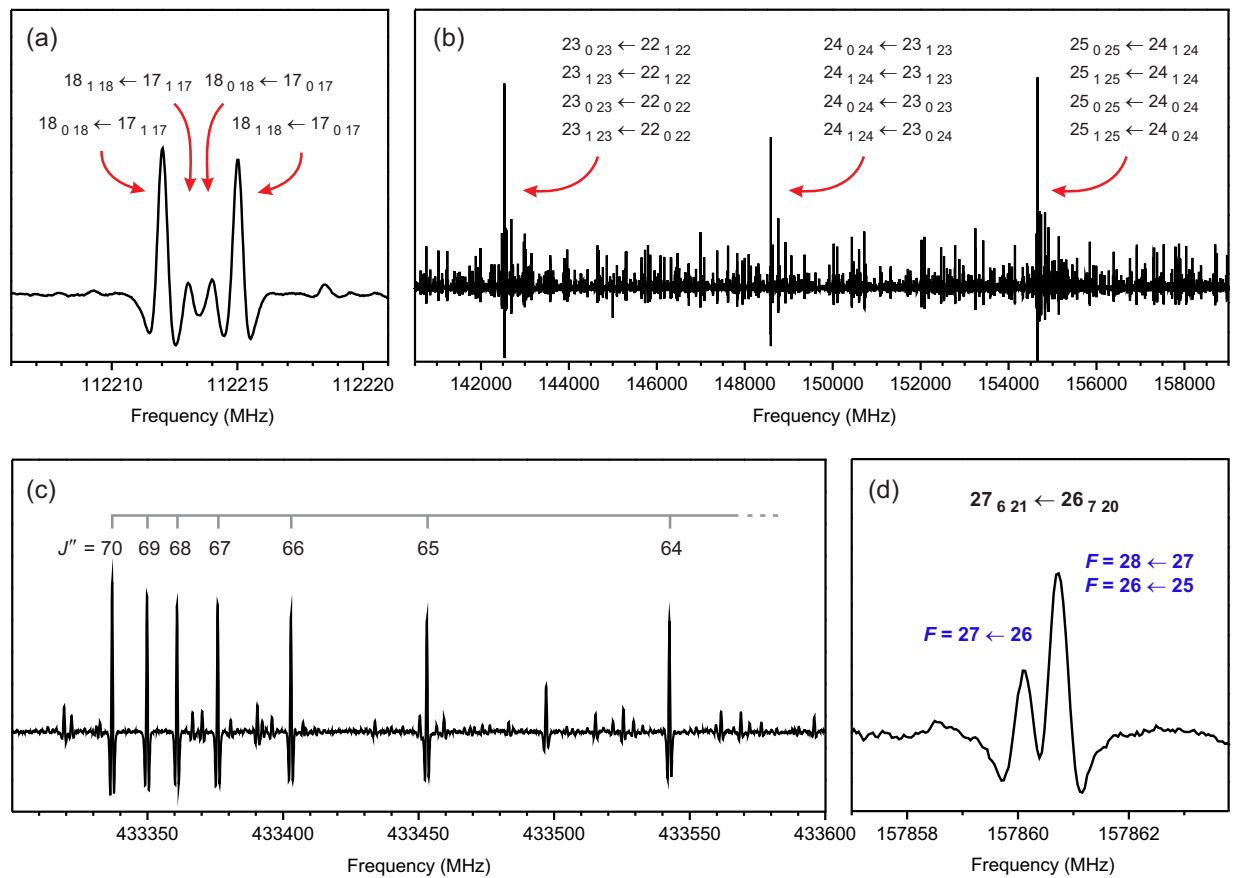}
      \caption{Illustrations of the main characteristics of the room-temperature rotational spectrum of propiolamide. (a) An example of a quartet of $a$-type and $b$-type R-branch rotational transitions between $K_{a}$ = 0, 1, $K_{c} = J$  pairs of energy levels. (b) Quadruply degenerate lines consisting of a pair of $a$-type and a pair of $b$-type transitions involving the same $K_{a}$ = 0, 1 pairs of levels. These strong lines appear in the spectrum approximately every 6.06 GHz, which corresponds to the value of $2C$. (c) An example of a group of high-$J$ quadruply degenerate transitions. The leading transition is $J = 71 \leftarrow 70$ and the value of $J$ decreases by 1 with each successive line running to higher frequencies. The $K_{a}$ quantum number is increasing away from the leading line in such a way that the first line corresponds to four degenerate $K_{a}= 0, 1$ transitions, the second to $K_{a}= 1, 2$, the third to $K_{a}= 2, 3$, and so on. (d) An example of $^{14}$N nuclear quadrupole hyperfine structure observed in the millimeter wave spectrum. The $F = J \pm 1$ hyperfine components typically appear overlapped, while the $F = J$ component is, in many cases, relatively well separated.}
      \label{features}
   \end{figure*}


\section{Experiments}
{\label{s:experiments}}

\begin{figure}[t]
   \centering
   \includegraphics[trim = 2mm 2mm 2mm 2mm, clip, width=8cm]{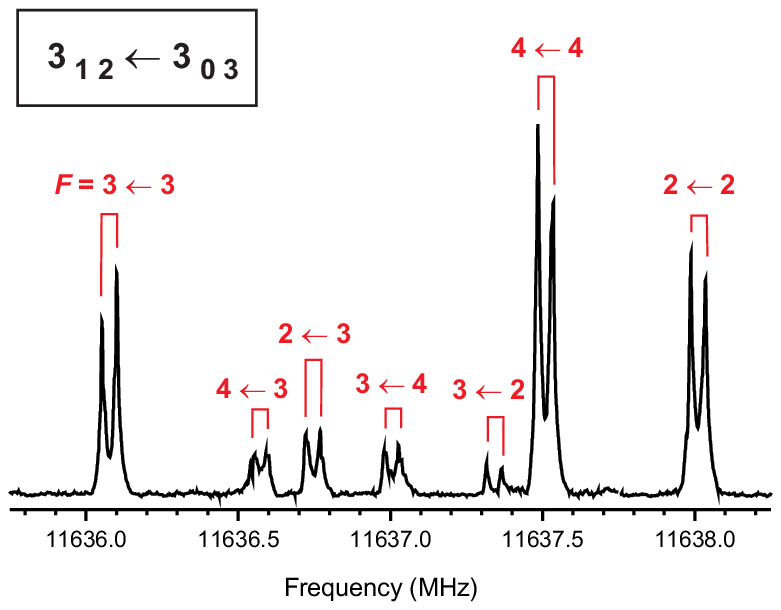}
      \caption{$^{14}$N nuclear quadrupole hyperfine structure of the $3_{1~2}\leftarrow 3_{0~3}$ rotational transition as measured by cavity-based FTMW spectrometer. Each hyperfine component, which appears as a Doppler doublet, is labeled with the corresponding values of $F$ quantum number.}\label{hfs}
   \end{figure}

Propiolamide was prepared from methyl propiolate, following the recipe of \citet{Miller1967}. The solid product was then evacuated overnight to remove traces of methanol and used without further purifications. To record the jet-cooled rotational
spectrum in the 6 – 18 GHz frequency range, propiolamide (melting point 57 -- 62$^{\circ}$C) was heated to approximately 80$^{\circ}$C in a pulsed nozzle, seeded in neon carrier gas (backing pressure of 1 bar) and adiabatically expanded into the Fabry-P\'{e}rot resonator of the FTMW spectrometer described elsewhere \citep{Alonso2015,Leon2017}. A short microwave radiation pulse of 0.3 ms duration was applied to polarize the molecules of propiolamide. A free induction decay associated with molecular emission was registered in the time domain and converted to the frequency domain by Fourier transformation. Since the molecular beam was directed along the resonator axis, the rotational transitions were observed as Doppler doublets (see Fig. \ref{hfs}). The resonance frequency was subsequently obtained as the arithmetic mean of the two Doppler components.

To record the room-temperature rotational spectra between 75 GHz and 440 GHz, propiolamide was placed into a small glass container that was connected directly to the free space cell of the millimeter wave spectrometer and evacuated.
The sample was heated with a heating gun (temperature of 80$^{\circ}$C of the heated air) until the pressure in the cell reached the optimal value between 10 and 20 $\mu$bar. The millimeter wave spectrometer employed in this work is based on sequential multiplication of the fundamental synthesizer frequency ($\leq$ 20 GHz) by a set of active and passive multipliers (VDI, Inc) and has been described elsewhere \citep{Kolesnikova2017a,Kolesnikova2017b}. For the measurements reported in this work, the synthesizer output was frequency modulated at a modulation frequency of 10.2 kHz and modulation depth between 20 kHz and 40 kHz before it was multiplied by amplifier-multiplier chains (WR10.0, WR6.5, WR9.0) in combination with additional doublers (WR4.3, WR2.2) and a tripler (WR2.8). After a double pass of the radiation through the cell, the signal was detected by zero-bias detectors and demodulated by a lock-in amplifier tuned to twice the modulation frequency. This demodulation procedure results in a shape of the lines that approximates the second derivative of a Gaussian line-profile function. All spectra were registered in 1~GHz sections in both directions (a single acquisition cycle) and averaged. These sections were ultimately combined into a single spectrum and further processed using the AABS package \citep{aabs,Kisiel2012}.

\section{Rotational spectra and analysis}
{\label{s:analysis}}

\subsection{Ground-state rotational spectrum}

As revealed by Stark modulation spectroscopy, propiolamide is essentially planar with dipole moment components $|\mu_{a}|=$ 1.08(2) D and $|\mu_{b}| =$ 3.51(2) D \citep{Little1978}. Both $a$-type and $b$-type transitions are therefore relevant in the millimeter wave spectrum. Strong lines corresponding to pairs of $b$-type transitions $(J+1)_{1,J+1}\leftarrow J_{0,J}$ and $(J+1)_{0,J+1}\leftarrow J_{1,J}$ could be immediately assigned in the spectrum. They were accompanied by weaker pairs of $a$-type transitions $(J+1)_{0,J+1}\leftarrow J_{0,J}$ and $(J+1)_{1,J+1}\leftarrow J_{1,J}$. Figure~\ref{features}a shows that the two $b$-type transitions straddle the pair of $a$-type transitions giving rise to characteristic quartets. These quartets, however, are not observed throughout the whole spectrum. As the $J$ quantum number increases, the participating energy levels become near degenerate and the four members of the quartets coalesce into very prominent quadruply degenerate lines that clearly dominate the spectrum shown in Fig. \ref{features}b. The same quartet structures and line blending are also observed for higher $K_{a}$ transitions. With the $J$ quantum number further progressively increasing, the quadruply degenerate lines with low values of $K_{a}$ quantum numbers start to form groups in which successive lines differ by a unit in $J$ and $K_{a}$. A typical example of this feature is shown in Fig. \ref{features}c. Such a pattern is characteristic for high-$J$ $a$-type and $b$-type spectra of planar and nearly planar molecules and was observed, for example, in acrylic acid \citep{Alonso2015}, phenol \citep{Kolesnikova2013}, or 2-chloroacrylonitrile \citep{Kisiel1994}. Finally, we also localized and measured $b$-type Q-branch transitions.

\begin{table*}
\caption{Spectroscopic constants of propiolamide in the ground state and three excited vibrational states ($A$-reduction, I$^{\text{r}}$-representation).}
\label{T1}
\begin{center}
\begin{footnotesize}
\setlength{\tabcolsep}{10pt}
\begin{tabular}{ l r r r r }
\hline\hline
 &   Ground state &  $v_{\text{in}}=1$  &  $v_{\text{out}}=1$  &  $v_{\text{inv}}=1$  \\
\hline
$A                   $  /               MHz     &  11417.92091 (11)\tablefootmark{a} &   11306.86167 (25)         &   11537.98892 (27)        &   11397.27733 (18)      \\
$B                   $  /               MHz     &   4135.477101 (40)                 &    4153.949250 (78)        &    4146.28849 (10)        &    4131.104876 (83)     \\
$C                   $  /               MHz     &   3032.592907(35)                  &    3036.029785 (61)        &    3041.801213 (72)       &    3032.360188 (56)     \\
$\Delta_{J}          $  /               kHz     &      0.579701 (18)                 &       0.612639 (36)        &       0.594171 (59)       &       0.578258 (51)     \\
$\Delta_{JK}         $  /               kHz     &     20.86142 (18)                  &      19.27979 (36)         &      21.82228 (41)        &      20.73217 (36)      \\
$\Delta_{K}          $  /               kHz     &    -10.05819 (20)                  &     -16.51167 (89)         &      -2.7538 (10)         &      -9.93128 (41)      \\
$\delta_{J}          $  /               kHz     &      0.1753084 (80)                &       0.190158 (18)        &       0.177915 (29)       &       0.174458 (25)     \\
$\delta_{K}          $  /               kHz     &     12.16107 (24)                  &      11.28587 (43)         &      13.08620 (65)        &      12.08963 (55)      \\
$\Phi_{J}            $  /                Hz     &   0.0003369 (31)                   &    0.0005094 (77)          &    0.000273 (17)          &    0.000370 (15)        \\
$\Phi_{JK}           $  /                Hz     &      0.19762 (12)                  &    0.16406 (25)            &    0.22129 (38)           &    0.19630 (28)         \\
$\Phi_{KJ}           $  /                Hz     &   -0.47001 (41)                    &    -0.55885 (90)           &    -0.3560 (13)           &    -0.46592 (78)        \\
$\Phi_{K}            $  /                Hz     &    0.30222 (35)                    &    -0.5968 (10)            &    1.1713 (13)            &    0.30279 (69)         \\
$\phi_{J}            $  /                Hz     &    0.0001448 (15)                  &    0.0002278 (41)          &    0.0001229 (89)         &    0.0001680 (78)       \\
$\phi_{JK}           $  /                Hz     &    0.095653 (64)                   &    0.08057 (11)            &    0.10591 (22)           &    0.09532 (19)         \\
$\phi_{K}            $  /                Hz     &    0.6385(11)                      &    0.3538 (18)             &   0.9125 (29)             &    0.6349 (22)          \\
$L_{JJK}             $  /               mHz     &     -0.0011188 (88)                &      -0.001301 (27)        &      -0.000844 (35)       &      -0.001004 (28)     \\
$L_{KKJ}             $  /               mHz     &   ...                              &       0.05208 (54)         &      -0.05083 (72)        &   ...                    \\
$\chi_{aa}           $  /               MHz     &   1.8157 (22)                      &   1.8157\tablefootmark{b}  &  1.8157\tablefootmark{b}  &   1.8157\tablefootmark{b} \\
$\chi_{bb}-\chi_{cc} $  /               MHz     &   6.0012 (44)                      &   6.336 (40)               &  6.212 (48)               &   5.796 (44)             \\
$J_{\text{min}}/J_{\text{max}}$                 &   0 / 75                           &   3 / 71                   &  3 / 71                   &   3 / 71                 \\
$K_{a\text{,min}}/K_{a\text{,max}}$             &   0 / 32                           &   0 / 25                   &  0 / 25                   &   0 / 25                 \\
$\sigma_{\text{fit}}$\tablefootmark{c}  MHz     &   0.023                            &   0.032                    &  0.032                    &   0.026                  \\
$N$\tablefootmark{d}                            &   1924                             &   1482                     &  1219                     &   978                    \\
\hline
\end{tabular}
\end{footnotesize}
\end{center}
\tablefoot{
\tablefoottext{a}{The numbers in parentheses are 1$\sigma$ uncertainties (67\% confidence level) in units of the last decimal digit. The SPFIT/SPCAT program package \citep{Pickett1991} was used for the analysis}
\tablefoottext{b}{Fixed to the ground state value owing to the limited hyperfine data set in excited vibrational states.}
\tablefoottext{c}{Root mean square deviation of the fit.}
\tablefoottext{d}{Number of distinct frequency lines in the fit.}}
\end{table*}

   \begin{figure*}[ht]
   \centering
   \includegraphics[trim = 10mm 5mm 10mm 5mm, clip, width=16.5cm]{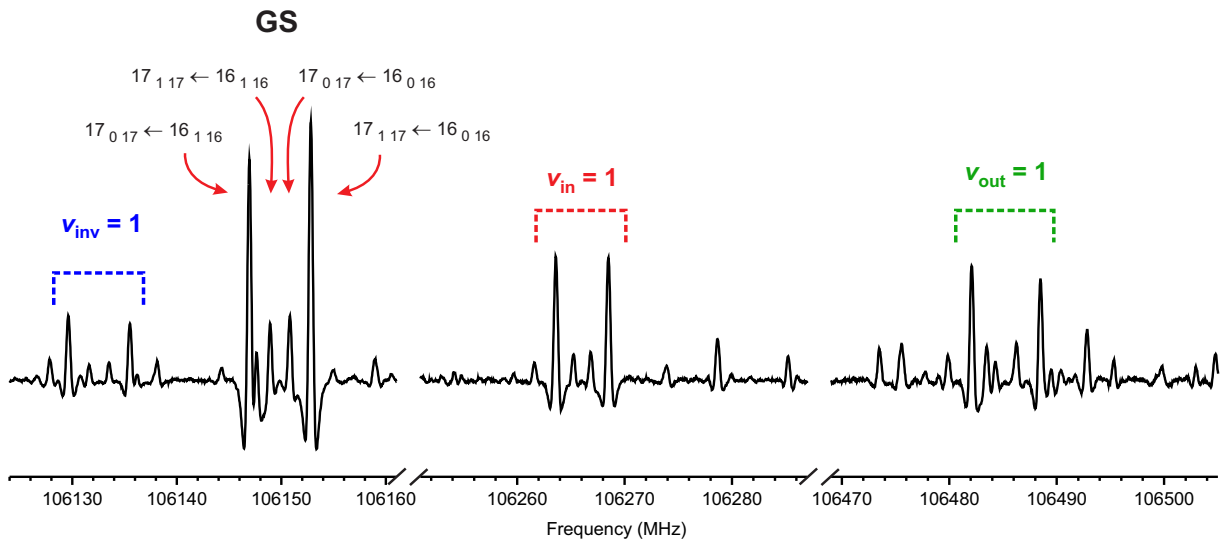}
      \caption{Vibrational satellites accompanying the quartet of ground-state $J = 17\leftarrow16$ transitions in the millimeter wave spectrum.}\label{excited}
   \end{figure*}

During the analysis procedure, we noticed that several transitions exhibited small splitting. We attributed this splitting to nuclear quadrupole coupling interactions of the single $^{14}$N nucleus present in the amide group. At the current resolution of the spectra in Fig.  \ref{features}, this typically gives rise to asymmetric doublets (see Fig. \ref{features}d). To correctly assign the hyperfine components blended together in these doublets, we measured and analyzed the nuclear quadrupole hyperfine structure from 6 GHz to 18 GHz using the cavity-based FTMW spectrometer that offers significantly higher spectral resolution. One of the measured transitions is exemplified in Fig. \ref{hfs}. In total 48 hyperfine components were measured and analyzed using the effective Hamiltonian $H=H_{\text{R}}+H_{\text{Q}}$, where $H_{\text{R}}$ is the standard Watson semi-rigid rotor Hamiltonian in $A$-reduction and I$^{\text{r}}$ representation \citep{Watson1977}, and $H_{\text{Q}}$ represents the nuclear quadrupole coupling Hamiltonian \citep{Gordy}. The coupling scheme $F=J+I_{\text{N}}$ between the rotational angular momentum $J$ and the $^{14}$N nuclear spin angular momentum $I_{\text{N}}$ was used. A comparison of the observed asymmetric doublets with predictions based on this hyperfine analysis made it possible to undertake a simultaneous analysis of hyperfine-split and hyperfine-free lines.
Transitions measured in the FTMW spectra and those in the millimeter wave spectra were adequately weighted assuming the experimental uncertainties of 5 and 30 kHz, respectively.
These data were ultimately combined with a selection of hyperfine-free transitions from \cite{Little1978} with assigned uncertainties equal to 100 kHz and were analyzed using the same effective Hamiltonian as mentioned above. The broad coverage of transition types and quantum number values (see Table \ref{T1}) guaranteed an accurate determination of rotational constants, centrifugal distortion constant along with the diagonal elements of the nuclear quadrupole coupling tensor. All these spectroscopic parameters are summarized in the first column of Table \ref{T1}. The list of measured transitions is given in Table \ref{S1} of the supplementary material.

\begin{table*}[h]
\caption{List of the measured transitions in the ground state and three excited vibrational states of propiolamide.}
\label{S1}
\begin{center}
\begin{footnotesize}
\setlength{\tabcolsep}{5pt}
\begin{tabular}{l r r r r r r r r r c c c }
\hline\hline
Vibrational state & $J'$ & $K_{a}'$  & $K_{c}'$ & $F'$ & $J''$ & $K_{a}''$ & $K_{c}''$ & $F''$ & $\nu_{\text{obs}}$ (MHz) \tablefootmark{a} &  $u_{\text{obs}}$ (MHz) \tablefootmark{b} & $\nu_{\text{obs}}-\nu_{\text{calc}}$ (MHz) \tablefootmark{c} & Comment \tablefootmark{d}\\
\hline
G.S.               &  2  &  0  &  2  &  1  &    1  &  1  &  1  &  1    &    6936.603  &  0.005   &  -0.001 &   (1)   \\
G.S.               &  5  &  0  &  5  & ... &    4  &  0  &  4  &  ...  &   33827.730  &  0.100   &  -0.036 &   (2)   \\
G.S.               & 28  &  5  & 23  & 28  &   28  &  4  & 24  &  28   &  109907.745  &  0.030   &  -0.025 &   (1)   \\
$v_{\text{in}}=1$  & 17  &  1  & 17  & ... &   16  &  0  & 16  & ...   &  106268.500  &  0.030   &  -0.015 &   (1)   \\
$v_{\text{in}}=1$  & 24  &  6  & 18  & 24  &   23  &  7  & 17  &  23   &  114610.591  &  0.030   &  -0.016 &   (1)   \\
$v_{\text{out}}=1$ & 19  &  1  & 19  & ... &   18  &  0  & 18  & ...   &  118650.237  &  0.030   &   0.015 &   (1)   \\
$v_{\text{out}}=1$ & 25  &  3  & 22  & 25  &   25  &  2  & 23  &  25   &  123309.311  &  0.030   &   0.006 &   (1)   \\
$v_{\text{inv}}=1$ & 16  &  2  & 15  & 16  &   16  &  1  & 16  &  16   &   89918.017  &  0.030   &   0.010 &   (1)   \\
$v_{\text{inv}}=1$ & 10  &  5  &  6  & ... &    9  &  4  &  5  & ...   &  141533.837  &  0.030   &   0.006 &   (1)   \\
\hline
\end{tabular}
\end{footnotesize}
\end{center}
\tablefoot{
\tablefoottext{a}{Observed frequency.}
\tablefoottext{b}{Uncertainty of the observed frequency.}
\tablefoottext{c}{Observed minus calculated frequency.}
\tablefoottext{d}{(1) This work. (2) Taken from \cite{Little1978}.}
This table is available in its entirety in electronic form at the CDS via anonymous ftp to cdsarc.u-strasbg.fr (130.79.128.5) or via
http://cdsweb.u-strasbg.fr/cgi-bin/qcat?J/A+A/. A portion is shown for guidance regarding its form and content.}
\end{table*}

\subsection{Rotational spectra in excited vibrational states}

Each ground-state line in the millimeter wave spectrum was accompanied by satellite lines at the high- and low-frequency sides. In the neighborhood of the ground-state $J = 17\leftarrow16$ transitions in Fig. 4, they appeared as quartets consisting of two $b$-type and two $a$-type transitions as in the ground state and were attributed to pure rotational transitions in excited vibrational states. According to our theoretical calculations at the B3LYP/6-311++G(d,p) level of the theory \citep[Gaussian 16 package,][]{g16} the three low-lying vibrational modes correspond mainly to the in-plane and out-of-plane bending motions of C$-$C$\equiv$C group $\nu_{\text{in}}$ ($A'$) and $\nu_{\text{out}}$ ($A''$), respectively, and to the NH$_{2}$ inversion $\nu_{\text{inv}}$ ($A''$).
Using the calculated first-order vibration-rotation constants $\alpha_{i}$, it was possible to predict the rotational constants for relevant excited states according to the equation $B_{v} = B_{e} - \sum_{i}\alpha_{i}(v_{i}+1/2)$, where $B_{v}$ and $B_{e}$ denote all the three rotational constants in a given excited state and in equilibrium, respectively, and $v_{i}$ is the vibrational quantum number of the $i$-th mode. The predictions proved to be sufficiently accurate for the unambiguous assignment of the satellite pattern, as shown in Fig. \ref{excited}. The two satellites displaced to high frequency from the ground state were assigned to the first quanta of the two bending modes $v_{\text{in}}=1$ and $v_{\text{out}}=1$, while the low-frequency satellite was ascribed to the first excited state of NH$_{2}$ inversion $v_{\text{inv}}=1$ (see Fig. \ref{excited}). Our assignments are in perfect agreement with those of \cite{Little1978}.

We performed the analysis of rotational transitions in the $v_{\text{in}}=1$, $v_{\text{out}}=1$, and $v_{\text{inv}}=1$ states in the same manner as for the ground state.\ The $a$-type and $b$-type R-branch transitions were identified and measured first. Then, we assigned $b$-type Q-branch transitions. As in the case of the ground state, several lines exhibited splitting due to $^{14}$N nuclear quadrupole coupling interactions. The above Hamiltonian $H=H_{\text{R}}+H_{\text{Q}}$ was therefore used to encompass all the assigned transitions. The obtained values of the rotational and centrifugal distortion constants as well as the $^{14}$N nuclear quadrupole coupling constants are collected in Table \ref{T1}. The list of the measured transition frequencies is given in Table \ref{S1}.

Although the rotational transitions in $v_{\text{in}}=1$, $v_{\text{out}}=1$, and $v_{\text{inv}}=1$ states from Table \ref{S1} were amenable to single state fits, departures of the centrifugal distortion constant values can be noticed for $v_{\text{in}}=1$ and $v_{\text{out}}=1$ when compared to the ground-state values. It is worth noting that the average value of $\Delta_{K}$ for these two states ($-9.63$ kHz) is very close to the ground-state value ($-10.058$ kHz). The same behavior is also perceptible for other centrifugal distortion constants and points to the existence of interactions between $v_{\text{in}}=1$ and $v_{\text{out}}=1$. The spectroscopic constants reported in Table \ref{T1} should be thus taken as effective parameters that reproduce the vibrational satellite spectrum near the experimental uncertainty.

\begin{table}
\caption{Partition functions of propiolamide.}
\label{T2}
\begin{center}
\begin{footnotesize}
\setlength{\tabcolsep}{15pt}
\begin{tabular}{ r r r r}
\hline\hline
$T$ (K) &  $Q_{\text{rot}}$ &  $Q_{\text{vib-rot}}\tablefootmark{a}$ & $Q_{\text{vib}}$ \\
\hline
   300.000    &   73387.03  &  149951.34  & 5.27 \\
   225.000    &   47641.04  &   83083.57  & 2.73 \\
   150.000    &   25920.14  &   35896.97  & 1.55 \\
    75.000    &    9161.75  &    9704.56  & 1.06 \\
    37.500    &    3240.31  &    3246.14  & 1.00 \\
    18.750    &    1146.94  &    1146.94  & 1.00 \\
     9.375    &     406.53  &     406.53  & 1.00 \\
\hline
\end{tabular}
\end{footnotesize}
\end{center}
\tablefoot{
\tablefoottext{a}{This vibrational-rotational partition function includes the contribution of the ground state and $v_{\text{in}}=1$, $v_{\text{out}}=1$, and $v_{\text{inv}}=1$ excited vibrational states.}}
\end{table}

Since the spectroscopic line lists for interstellar detection require not only reliable
line frequencies but also line intensities, we provide in Table \ref{T2} partition functions of propiolamide at multiple temperatures. The rotational partition function ($Q_{\text{rot}}$) corresponds to the ground vibrational state and was evaluated by summation over the energy levels. We used the SPCAT program \citep{Pickett1991} to undertake this summation numerically, employing the rotational and centrifugal distortion constants from Table \ref{T1}, dipole moment components $|\mu_{a}|=$ 1.08 D and $|\mu_{b}| =$ 3.51 D from \cite{Little1978}, and the maximum value of the $J$ quantum number of 200.
The vibrational-rotational partition function ($Q_{\text{vib-rot}}$) takes into consideration the ground state and the three observed excited vibrational states. This partition function was also computed by summation over the energy levels with the vibrational energies of 170, 210, and 300~cm$^{-1}$ for $v_{\text{in}}=1$, $v_{\text{out}}=1$, and $v_{\text{inv}}=1$, respectively, and the rotational and centrifugal distortion constants from Table \ref{T1}. The values of the vibrational energies were roughly estimated by comparison of spectral intensities of the same rotational transitions in the corresponding excited vibrational state with respect to those in the ground state.
Finally, the vibrational partition function ($Q_{\text{vib}}$) was evaluated using Eq. 3.60 of \cite{Gordy1970} by taking into consideration the frequencies of 18 normal vibrational modes from Table~\ref{freq} of the Appendix.

\begin{figure*}[!t]
\centerline{\resizebox{0.88\hsize}{!}{\includegraphics[angle=0]{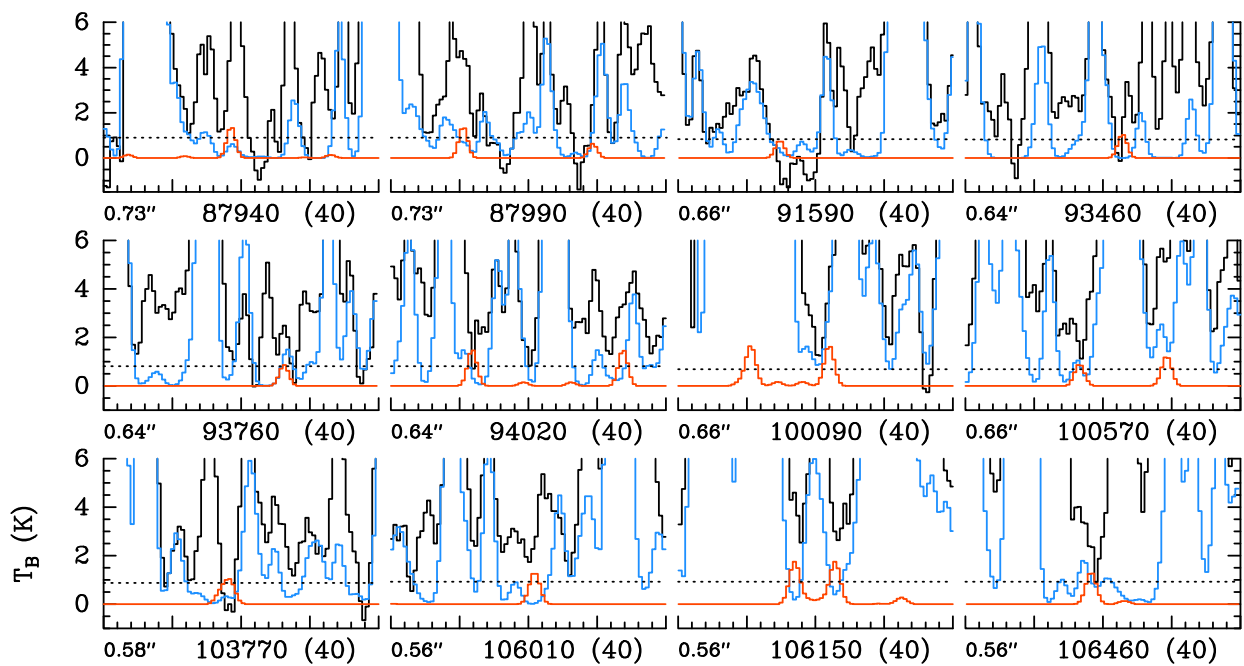}}}
\caption{Selection of transitions of HCCC(O)NH$_2$, $\varv = 0$ covered by our
ALMA survey. The synthetic spectrum of HCCC(O)NH$_2$, $\varv = 0$ used to
derive the upper limit to its column density is indicated in red and overlaid
on the observed spectrum of Sgr~B2(N1S) shown in black. The blue synthetic
spectrum contains the contributions from all molecules identified in our survey
so far, but not from the species shown in red.
The central frequency and width are indicated in MHz below each panel. The
angular resolution (HPBW) is also indicated. The $y$-axis is labeled in
brightness temperature units (K). The dotted line indicates the $3\sigma$
noise level.}
\label{f:spec_hccconh2_ve0_n1s}
\end{figure*}

\begin{table*}[!ht]
 \begin{center}
 \caption{
 Parameters of our best-fit LTE model of acetamide toward Sgr~B2(N1S), and upper limit for propiolamide.
}
 \label{t:coldens_n1s}
 \vspace*{-1.2ex}
 \begin{tabular}{lcrccccccr}
 \hline\hline
 \multicolumn{1}{c}{Molecule} & \multicolumn{1}{c}{Status\tablefootmark{a}} & \multicolumn{1}{c}{$N_{\rm det}$\tablefootmark{b}} & \multicolumn{1}{c}{$\theta_{\rm s}$\tablefootmark{c}} & \multicolumn{1}{c}{$T_{\mathrm{rot}}$\tablefootmark{d}} & \multicolumn{1}{c}{$N$\tablefootmark{e}} & \multicolumn{1}{c}{$F_{\rm vib}$\tablefootmark{f}} & \multicolumn{1}{c}{$\Delta V$\tablefootmark{g}} & \multicolumn{1}{c}{$V_{\mathrm{off}}$\tablefootmark{h}} & \multicolumn{1}{c}{$\frac{N_{\rm ref}}{N}$\tablefootmark{i}} \\ 
  & & & \multicolumn{1}{c}{\small ($''$)} & \multicolumn{1}{c}{\small (K)} & \multicolumn{1}{c}{\small (cm$^{-2}$)} & & \multicolumn{1}{c}{\small (km~s$^{-1}$)} & \multicolumn{1}{c}{\small (km~s$^{-1}$)} & \\ 
 \hline
 CH$_3$C(O)NH$_2$\tablefootmark{(j)}$^\star$ & d & 153 &  2.0 &  160 &  4.1 (17) & 1.16 & 5.0 & $0.0$ &       1 \\ 
\hline 
 HCCC(O)NH$_2$ & n & 0 &  2.0 &  160 & $<$  8.5 (15) & 1.18 & 5.0 & $0.0$ & $>$      48 \\ 
\hline 
 \end{tabular}
 \end{center}
 \vspace*{-2.5ex}
 \tablefoot{
 \tablefoottext{a}{d: detection, n: nondetection.}
 \tablefoottext{b}{Number of detected lines \citep[conservative estimate, see Sect.~3 of][]{Belloche16}. One line of a given species may mean a group of transitions of that species that are blended together.}
 \tablefoottext{c}{Source diameter (FWHM).}
 \tablefoottext{d}{Rotational temperature.}
 \tablefoottext{e}{Total column density of the molecule. $x$ ($y$) means $x \times 10^y$.}
 \tablefoottext{f}{Correction factor that was applied to the column density to account for the contribution of vibrationally excited states, in the cases where this contribution was not included in the partition function of the spectroscopic predictions.}
 \tablefoottext{g}{Linewidth (FWHM).}
 \tablefoottext{h}{Velocity offset with respect to the assumed systemic velocity of Sgr~B2(N1S), $V_{\mathrm{sys}} = 62$ km~s$^{-1}$.}
 \tablefoottext{i}{Column density ratio, with $N_{\rm ref}$ the column density of the previous reference species marked with a $\star$.}
 \tablefoottext{j}{The parameters were derived from the ReMoCA survey by \citet{Belloche19}.}
 }
 \end{table*}

\begin{table*}[!ht]
 \begin{center}
 \caption{
 Parameters of our best-fit LTE model of acetamide toward Sgr~B2(N2), and upper limit for propiolamide.
}
 \label{t:coldens_n2}
 \vspace*{-1.2ex}
 \begin{tabular}{lcrccccccr}
 \hline\hline
 \multicolumn{1}{c}{Molecule} & \multicolumn{1}{c}{Status\tablefootmark{a}} & \multicolumn{1}{c}{$N_{\rm det}$\tablefootmark{b}} & \multicolumn{1}{c}{$\theta_{\rm s}$\tablefootmark{c}} & \multicolumn{1}{c}{$T_{\mathrm{rot}}$\tablefootmark{d}} & \multicolumn{1}{c}{$N$\tablefootmark{e}} & \multicolumn{1}{c}{$F_{\rm vib}$\tablefootmark{f}} & \multicolumn{1}{c}{$\Delta V$\tablefootmark{g}} & \multicolumn{1}{c}{$V_{\mathrm{off}}$\tablefootmark{h}} & \multicolumn{1}{c}{$\frac{N_{\rm ref}}{N}$\tablefootmark{i}} \\ 
  & & & \multicolumn{1}{c}{\small ($''$)} & \multicolumn{1}{c}{\small (K)} & \multicolumn{1}{c}{\small (cm$^{-2}$)} & & \multicolumn{1}{c}{\small (km~s$^{-1}$)} & \multicolumn{1}{c}{\small (km~s$^{-1}$)} & \\ 
 \hline
 CH$_3$C(O)NH$_2$\tablefootmark{(j)}$^\star$ & d & 23 &  0.9 &  180 &  1.4 (17) & 1.23 & 5.0 & $1.5$ &       1 \\ 
\hline 
 HCCC(O)NH$_2$ & n & 0 &  0.9 &  180 & $<$  1.0 (16) & 1.30 & 5.0 & $0.0$ & $>$      13 \\ 
\hline 
 \end{tabular}
 \end{center}
 \vspace*{-2.5ex}
 \tablefoot{
 \tablefoottext{a}{d: detection, n: nondetection.}
 \tablefoottext{b}{Number of detected lines \citep[conservative estimate, see Sect.~3 of][]{Belloche16}. One line of a given species may mean a group of transitions of that species that are blended together.}
 \tablefoottext{c}{Source diameter (FWHM).}
 \tablefoottext{d}{Rotational temperature.}
 \tablefoottext{e}{Total column density of the molecule. $x$ ($y$) means $x \times 10^y$.}
 \tablefoottext{f}{Correction factor that was applied to the column density to account for the contribution of vibrationally excited states, in the cases where this contribution was not included in the partition function of the spectroscopic predictions.}
 \tablefoottext{g}{Linewidth (FWHM).}
 \tablefoottext{h}{Velocity offset with respect to the assumed systemic velocity of Sgr~B2(N2), $V_{\mathrm{sys}} = 74$ km~s$^{-1}$.}
 \tablefoottext{i}{Column density ratio, with $N_{\rm ref}$ the column density of the previous reference species marked with a $\star$.}
 \tablefoottext{j}{The parameters were derived from the EMoCA survey by \citet{Belloche17}.}
 }
 \end{table*}

\section{Search for propiolamide toward Sgr~B2(N)}
\label{s:astro}

\subsection{Observations}
\label{ss:observations}

The data set used in this work was extracted from the ReMoCA survey performed with ALMA toward Sgr~B2(N). Details
about the observational setup and data reduction of this imaging spectral line survey can be found in
\citet{Belloche19}. In short, the angular resolution (HPBW) of the survey
varies between $\sim$0.3$\arcsec$ and $\sim$0.8$\arcsec$, with a median value
of 0.6$\arcsec$, which corresponds to $\sim$4900~au at the distance of Sgr~B2
\citep[8.2~kpc;][]{Reid19}. The observations covered the frequency range from
84.1~GHz to 114.4~GHz with a spectral resolution of 488~kHz (1.7 to
1.3~km~s$^{-1}$). The equatorial coordinates of the phase center are
($\alpha, \delta$)$_{\rm J2000}$=
($17^{\rm h}47^{\rm m}19{\fs}87, -28^\circ22'16{\farcs}0$). This position is
located halfway between the two main hot molecular cores of Sgr~B2(N) called
Sgr~B2(N1) and Sgr~B2(N2), which are separated by 4.9$\arcsec$ or $\sim$0.2~pc
in projection onto the plane of the sky. The sensitivity per spectral channel
ranges between 0.35~mJy~beam$^{-1}$ and 1.1~mJy~beam$^{-1}$ (rms), with a median
value of 0.8~mJy~beam$^{-1}$.

We selected two positions for this study: the offset position Sgr~B2(N1S)
located at ($\alpha, \delta$)$_{\rm J2000}$=
($17^{\rm h}47^{\rm m}19{\fs}870$, $-28^\circ22\arcmin19{\farcs}48$) and the
secondary hot core Sgr~B2(N2) at ($\alpha, \delta$)$_{\rm J2000}$=
($17^{\rm h}47^{\rm m}19{\fs}863$, $-28^\circ22\arcmin13{\farcs}27$). Sgr~B2(N1S)
is about 1$\arcsec$ to the south of the main hot core Sgr~B2(N1) and was
chosen by \citet{Belloche19} for the lower opacity of its continuum emission,
which allows for a deeper look into the molecular content of this hot core. We
used a new version of our data set for which we have improved the splitting of
the continuum and line emission as reported in \citet{Melosso20}.

The spectra of Sgr~B2(N1S) and Sgr~B2(N2) were modeled with the software Weeds
\citep[][]{Maret11} under the assumption of local thermodynamic equilibrium
(LTE). This assumption is appropriate because the regions where hot-core
emission is detected in Sgr~B2(N) have high densities
\citep[$>1 \times 10^{7}$~cm$^{-3}$, see][]{Bonfand19}. We derived a best-fit
synthetic spectrum of each molecule separately and then added the
contributions of all identified molecules together. We modeled each species
with a set of five parameters: size of the emitting region ($\theta_{\rm s}$),
column density ($N$), temperature ($T_{\rm rot}$), linewidth ($\Delta V$), and
velocity offset ($V_{\rm off}$) with respect to the assumed systemic velocity
of the source, $V_{\rm sys}=62$~km~s$^{-1}$ for Sgr~B2(N1S) and
$V_{\rm sys}= 74$~km~s$^{-1}$ for Sgr~B2(N2).

\begin{figure*}[!t]
\centerline{\resizebox{0.88\hsize}{!}{\includegraphics[angle=0]{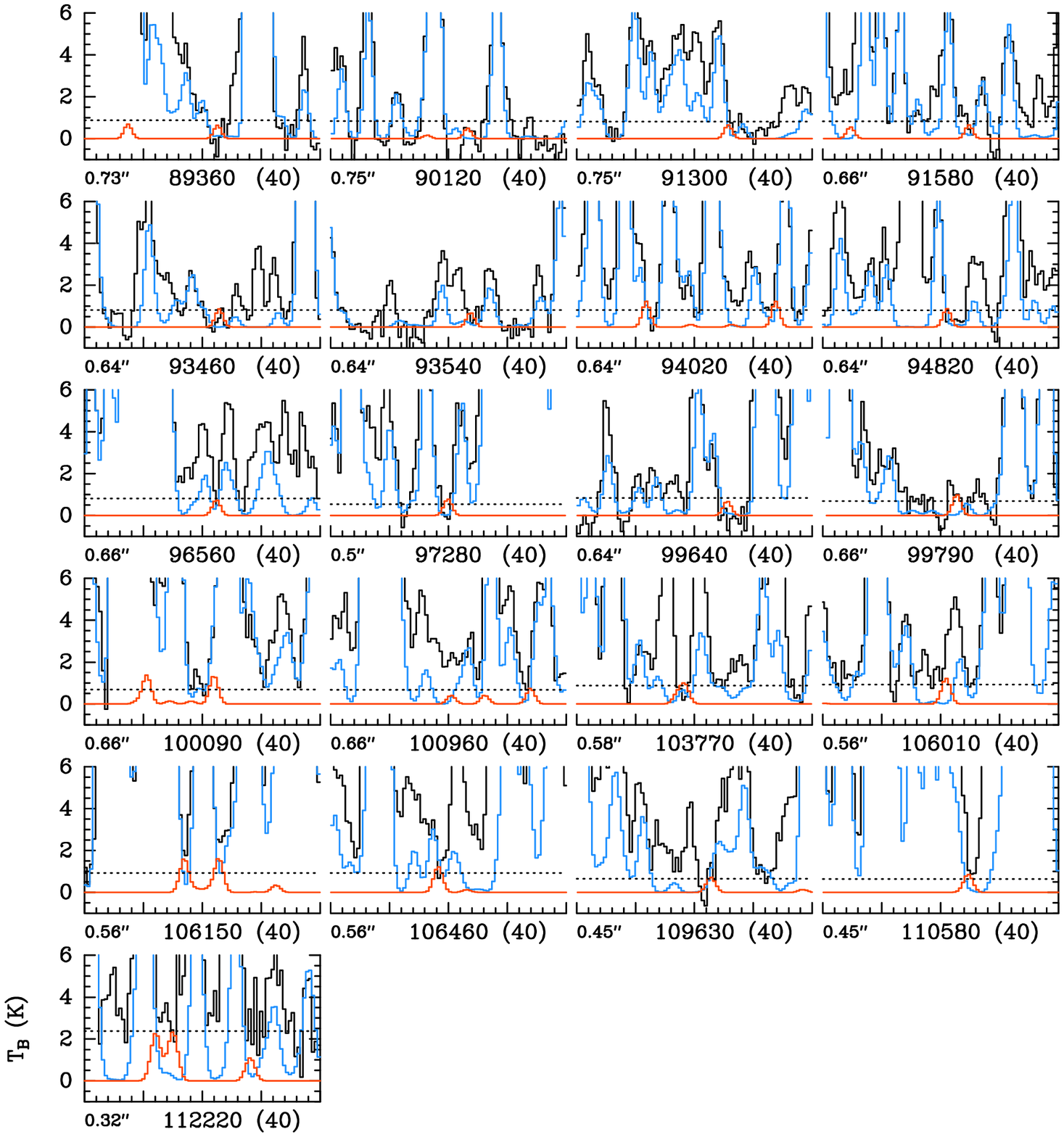}}}
\caption{Same as Fig.~\ref{f:spec_hccconh2_ve0_n1s} but for Sgr~B2(N2).}
\label{f:spec_hccconh2_ve0_n2}
\end{figure*}

\subsection{Nondetection of propiolamide}
\label{ss:nondetection}

We used the parameters derived for acetamide, CH$_3$C(O)NH$_2$, by
\citet{Belloche17,Belloche19} to compute LTE synthetic spectra of propiolamide
and search for rotational emission of the latter in the ReMoCA spectra of
Sgr~B2(N2) and Sgr~B2(N1S), respectively. We only kept the column density of
propiolamide as a free parameter. We found no evidence for emission of
propiolamide toward either source. The nondetection toward Sgr~B2(N1S)
and Sgr~B2(N2) is illustrated in Figs.~\ref{f:spec_hccconh2_ve0_n1s} and
\ref{f:spec_hccconh2_ve0_n2}, respectively, and the upper limits to the column
density of propiolamide are indicated in Tables~\ref{t:coldens_n1s} and
\ref{t:coldens_n2}, respectively. The tables also recall the parameters
previously obtained for acetamide. The spectroscopic catalog used to
compute the synthetic spectra of propiolamide shown in
Figs.~\ref{f:spec_hccconh2_ve0_n1s} and \ref{f:spec_hccconh2_ve0_n2} was
prepared with the partition function that includes the vibrational ground
state and the three lowest vibrationally excited states ($Q_{\text{vib-rot}}$ in Table \ref{T2}). The correction factor
$F_{\rm vib}$ used in Tables~\ref{t:coldens_n1s} and \ref{t:coldens_n2} to
derive the upper limit to the column density of propiolamide accounts for the
higher vibrational states, using the corresponding vibrational frequencies
listed in Table~\ref{freq}. $F_{\rm vib}$ was computed as
$Q_{\text{vib}}\times Q_{\text{rot}}/Q_{\text{vib-rot}}$ using the values listed in Table \ref{T2}.

\section{Discussion}
\label{s:discussion}

\subsection{Comparison to other molecules in Sgr~B2(N)}
\label{ss:discussion_astro}

The nondetection of propiolamide toward Sgr~B2(N1S) and Sgr~B2(N2) reported
in Sect.~\ref{ss:nondetection} implies that propiolamide is at least
$\sim$50 and $\sim$13 times less abundant than acetamide toward Sgr~B2(N1S)
and Sgr~B2(N2), respectively. For comparison, cyanoacetylene, HC$_3$N, is
about six times less abundant than methyl cyanide, CH$_3$CN, toward Sgr~B2(N2)
\citep[][]{Belloche16} and a preliminary analysis of the ReMoCA survey yields
a similar ratio for Sgr~B2(N1S). Both pairs of molecules, the amides
HCCC(O)NH$_2$/CH$_3$C(O)NH$_2$ and the cyanides HC$_3$N/CH$_3$CN, share the
same structural difference (an unsaturated C$_2$H group replaced with a
saturated CH$_3$ group), but the abundance difference between both (larger)
amides is more pronunced than for the (smaller) cyanides, by at least a factor
of 2 for Sgr~B2(N2) and at least a factor of 8 for Sgr~B2(N1S).

\subsection{Formation mechanisms for propiolamide}
\label{ss:discussion_chemistry}

Existing interstellar chemical networks do not include propiolamide; however, guided by the behavior of related species in the chemical kinetic simulations, we may speculate on the factors that could impact upon its formation and/or survival during star formation.

In common with acetamide, it would appear plausible that propiolamide could be produced on grain surfaces, or within their ice mantles, by the reaction of the radical NH$_2$CO with another radical; CH$_3$ in the former case, C$_2$H in the latter. Unsaturated hydrocarbons such as acetylene (C$_2$H$_2$) and associated radicals such as C$_2$H, are typically considered ``early-time'' species, which become abundant in the gas phase before free carbon has had time to become locked up in stable species such as CO. At such a stage, the dust-grain ice mantles are still in the process of formation. Surface radical reactions during this period might provide an opportunity for propiolamide production, while gas-phase hydrocarbons remain available to adsorb onto the grain surfaces. Although the grains would be considered too cold ($\sim$10~K) for surface diffusion of large radicals to drive production, the occasional presence of reactive radicals in close proximity to each other appears to be sufficient to allow a degree of complex organic molecule production even at low temperatures \citep[e.g.,][]{Jin20}. New models of hot-core chemistry (Garrod et al., in prep.) indicate that such cold, early mechanisms may be major contributors to the production of certain species that are later observed in the gas phase. This appears to be the case with acetamide; if similar cold mechanisms were also to contribute to propiolamide production, then the ratio of methane (CH$_4$) to acetylene produced in the models could give an indication of the relative production of the larger species. At the end of the cold collapse stage, the Garrod et al. (in prep.) models indicate that methane is around 3 orders of magnitude more abundant than acetylene in the ices. This ratio is consistent with the observational ratio of acetamide to the upper limit for propiolamide.

In the models, acetamide may also be formed on the grains through the addition of the radicals NH$_2$ and CH$_3$CO, the latter of which is a product of the destruction of acetaldehyde (CH$_3$CHO). The comparable mechanism for propiolamide production would be NH$_2$ addition to the HC$_2$CO radical, which could be formed by the loss of a hydrogen atom from the aldehyde group of propynal, HC$_2$CHO. As this molecule and its surrounding chemistry are not included in our chemical models (other than in the most trivial fashion), the implications of such a mechanism cannot be quantified.

An alternative to cold surface production at early times is the cosmic-ray-induced UV photolysis of molecules within the dust-grain ice mantles, which may occur continuously from the onset of mantle formation until their ultimate desorption into the gas phase. This mechanism provides a means by which the requisite radicals can be produced and thence react to form propiolamide and acetamide. In this case, the relative abundances of precursors NH$_2$CHO, CH$_4$ (and/or CH$_3$OH) and C$_2$H$_2$, as well as NH$_3$, CH$_3$CHO and HC$_2$CHO, would again be important in determining ratios between acetamide and propiolamide.

Assuming that propiolamide may be formed in some abundance on dust grains prior to the thermal desorption of dust-grain ices, it is also possible that reactions with abundant grain-surface H atoms could produce at least partial conversion to propanamide (CH$_3$CH$_2$CONH$_2$), further reducing the abundance of propiolamide. Hydrogenation by H atoms could similarly act to reduce the abundance of acetylene in the ice mantles, removing a possible precursor for the ongoing production of propiolamide via ice photolysis.
All in all, the observed abundances and ratios between propiolamide and acetamide appear to be consistent with the expectations from limited chemical modeling evidence.

\section{Conclusions}
\textbf{\label{s:conclusions}}

Using a combination of time-domain FTMW spectroscopy and frequency-domain millimeter wave spectroscopy techniques, a detailed rotational study of propiolamide from 6 GHz to 440 GHz was carried out. In total, more than 5500 new rotational lines for the ground state and the three lowest excited vibrational states were measured and assigned. The present work significantly extends the frequency coverage of the propiolamide rotational spectrum known to date and newly derived spectroscopic parameters provide a firm base to guide a search for this molecule using radio astronomy.

Propiolamide was not detected toward the hot molecular cores Sgr~B2(N1S)
and Sgr~B2(N2) with ALMA. The upper limits derived for its column density
imply that it is at least 50 and 13 times less abundant than
acetamide toward these sources, respectively. This abundance
difference between both amides is more pronounced than for their corresponding
nitriles by at least a factor of 8 for Sgr~B2(N1S) and 2 for Sgr~B2(N2).

The observational results seem consistent with the low ratio of
propiolamide to acetamide inferred from the results of chemical kinetic models
(which include the latter species but not the former). In the proposed
scenario, production of propiolamide would occur through radical addition on
dust-grain surfaces. This mechanism may be most effective at early times in
the chemical evolution when unsaturated hydrocarbons should be abundant in
the gas phase. Reactions with grain-surface H atoms could also diminish the
abundance of propiolamide and its precursors in the ice mantles prior to
thermal desorption.

More promising sources for the detection of propiolamide may be
sources where unsaturated carbon chain molecules are more prominent
than in Sgr B2(N). One possibility, for instance, is TMC1, where HC$_3$N is
more than one order of magnitude more abundant than CH$_3$CN
\citep{Gratier2016}. However, the amine CH$_3$NH$_2$ has not been
detected toward TMC1 in the GOTHAM survey with the GBT telescope so
far (B. McGuire, priv. comm. 2021), which is probably not a good
sign for propiolamide. The quiescent source G+0.693-0.027, which is
located in the vicinity of Sgr B2(N) and where propargylimine,
HCCCHNH, was recently detected by \cite{Bizzocchi2020}, may be a
more promising source to continue the search for propiolamide.

\begin{acknowledgements}

The funding from Ministerio de Ciencia e Innovaci\'{o}n (CTQ2016-76393-P
and PID2019-111396GB-I00), Czech Science Foundation (GACR, grant
19-25116Y), Junta de Castilla y Le\'{ó}n (Grants VA077U16 and VA244P20), and
European Research Council under the European Union’s Seventh Framework
Programme ERC-2013-SyG, Grant Agreement n. 610256 NANOCOSMOS are
gratefully acknowledged. E.R.A. thanks Fundaci\'{o}n Biof\'{\i}sica Bizkaia (FBB)
for a postdoctoral contract.
J.C.G. thanks the program Physique et Chimie du Milieu Interstellaire (INSU-CNRS) and the Centre National d’Etudes Spatiales (CNES).
This paper makes use of the following ALMA data:
ADS/JAO.ALMA\#2016.1.00074.S.
ALMA is a partnership of ESO (representing its member states), NSF (USA), and
NINS (Japan), together with NRC (Canada), NSC and ASIAA (Taiwan), and KASI
(Republic of Korea), in cooperation with the Republic of Chile. The Joint ALMA
Observatory is operated by ESO, AUI/NRAO, and NAOJ. The interferometric data
are available in the ALMA archive at https://almascience.eso.org/aq/.
Part of this work has been carried out within the Collaborative
Research Centre 956, sub-project B3, funded by the Deutsche
Forschungsgemeinschaft (DFG) -- project ID 184018867.
RTG acknowledges support from the National Science Foundation (grant No.
AST 19-06489).

\end{acknowledgements}



\bibliographystyle{aa}
\bibliography{library}



\begin{appendix}

\section{Complementary Tables}

\begin{table*}
\caption{Frequencies of normal vibrational modes of propiolamide used to calculate the vibrational partition function.}
\label{freq}
\begin{center}
\begin{footnotesize}
\setlength{\tabcolsep}{4pt}
\begin{tabular}{r r r }
\hline\hline
Mode & Frequency (cm$^{-1}$) & Symmetry \\
\hline
 1    &  3724   &   $A^\prime$     \\
 2    &  3588   &   $A^\prime$     \\
 3    &  3471   &   $A^\prime$     \\
 4    &  2209   &   $A^\prime$     \\
 5    &  1747   &   $A^\prime$     \\
 6    &  1614   &   $A^\prime$     \\
 7    &  1329   &   $A^\prime$     \\
 8    &  1090   &   $A^\prime$     \\
 9    &   797   &   $A^\prime$     \\
10    &   701   &   $A^\prime$     \\
11    &   600   &   $A^\prime$     \\
12    &   490   &   $A^\prime$     \\
13    &   170   &   $A^\prime$     \\
14    &   762   &   $A^{\prime\prime}$  \\
15    &   723   &   $A^{\prime\prime}$  \\
16    &   554   &   $A^{\prime\prime}$  \\
17    &   300   &   $A^{\prime\prime}$  \\
18    &   230   &   $A^{\prime\prime}$  \\
\hline
\end{tabular}
\end{footnotesize}
\end{center}
\tablefoot{The frequencies of the three lowest modes were estimated on the basis of the experimental relative intensities measurements with uncertainties of 30 cm$^{-1}$, while the others were taken from the theoretical calculations at the B3LYP/6-311++G(d,p) level of the theory \citep[Gaussian 16 package,][]{g16}.}
\end{table*}

\end{appendix}

\end{document}